\documentclass[]{aastex631}

\begin{document}

\title{Detecting quasi-periodic pulsations in solar and stellar flares with a neural network}

\correspondingauthor{Sergey A. Belov}
\email{Sergey.Belov@warwick.ac.uk}

\author[0000-0002-3505-9542]{Sergey A. Belov}
\affiliation{Centre for Fusion, Space and Astrophysics, Department of Physics, University of Warwick, Coventry CV4 7AL, UK}

\author[0000-0002-0687-6172]{Dmitrii Y. Kolotkov}
\affiliation{Centre for Fusion, Space and Astrophysics, Department of Physics, University of Warwick, Coventry CV4 7AL, UK}
\affiliation{Engineering Research Institute \lq\lq Ventspils International Radio Astronomy Centre (VIRAC) \rq\rq, Ventspils University of Applied Sciences, Ventspils, LV-3601, Latvia}

\author[0000-0001-6423-8286]{Valery M. Nakariakov}
\affiliation{Centre for Fusion, Space and Astrophysics, Department of Physics, University of Warwick, Coventry CV4 7AL, UK}
\affiliation{Engineering Research Institute \lq\lq Ventspils International Radio Astronomy Centre (VIRAC) \rq\rq, Ventspils University of Applied Sciences, Ventspils, LV-3601, Latvia}

\author[0000-0002-5209-9378]{Anne-Marie Broomhall}
\affiliation{Centre for Fusion, Space and Astrophysics, Department of Physics, University of Warwick, Coventry CV4 7AL, UK}

\begin{abstract}

Quasi-periodic pulsations (QPP) are often detected in solar and stellar flare lightcurves. These events may contain valuable information about the underlying fundamental plasma dynamics as they are not described by the standard flare model. The detection of QPP signals in flare lightcurves is hindered by their intrinsically non-stationary nature, contamination by noise, and the continuously increasing amount of flare observations. Hence, the creation of automated techniques for QPP detection is imperative. 
We implemented the Fully Convolution Network (FCN) architecture to classify the flare lightcurves whether they have exponentially decaying harmonic QPP or not.  To train the FCN, 90,000 synthetic flare lightcurves with and without QPP were generated. After training, it showed an accuracy of 87.2\% on the synthetic test data and did not experience overfitting.
To test the FCN performance on real data, we used the subset of stellar flare lightcurves observed by Kepler, with strong evidence of decaying QPP identified hitherto with other methods.
Then, the FCN was applied to find QPPs in a larger-scale Kepler flare catalogue comprised of 2274 events, resulting in a 7\% QPP detection rate with a probability above 95\%.
The FCN, implemented in Python, is accessible through a browser application with a user-friendly graphical interface and detailed installation and usage guide.
The obtained results demonstrate that the developed FCN performs well and successfully detects exponentially decaying harmonic QPP in real flare data, and can be used as a tool for preliminary sifting of the QPP events of this type in future large-scale observational surveys.

\end{abstract}

\keywords{Sun: flares --- Sun: oscillations --- stars: flare --- techniques: miscellaneous}

\section{Introduction} \label{sec:intro}

Solar flares, together with coronal mass ejections, are the most powerful physical processes in the solar system \citep[e.g.,][]{2017LRSP...14....2B}.
Understanding the physical processes which cause the impulsive releases of the magnetic energy in the solar atmosphere are among the key challenges of modern solar physics. Furthermore, there is a growing interest in similar phenomena detected on other stars, including the sun-like stars with potentially habitable planetary systems.
The so-called standard model of a solar flare attributes the energy release to the process of magnetic reconnection \citep[e.g.,][]{2011LRSP....8....6S}. 

An intensively studied phenomenon which is not described by the standard flare model are quasi-periodic pulsations (QPP) of the flaring emission \citep[e.g.,][]{2009SSRv..149..119N, 2010PPCF...52l4009N, 2021SSRv..217...66Z}. QPP appear as the quasi-periodic modulation of the flaring emission in all observational bands, from radio to gamma-rays, in both thermal and non-thermal emission, and in flares of all classes, from microflares \citep[e.g.,][]{2018ApJ...859..154N} to the most-power flare classes \citep[e.g.,][]{2018ApJ...858L...3K}. Typical periods of QPP range from a fraction of a second to several minutes. \citet{2015SoPh..290.3625S} demonstrated that about 80\% of X-class solar flares display QPP in the soft X-ray emission. Furthermore, \citet{2020ApJ...895...50H}
found out that approximately 46\% of X-class, 29\% of M-class, and 7\% of C-class flares show evidence of stationary, i.e., narrowband QPP. Those estimations are highly sensitive to the detection criterion, i.e., upon the definition of a QPP. One of the complications is the essentially non-stationary nature of QPP \citep{2019PPCF...61a4024N, 2020ApJ...905...70P, 2023MNRAS.523.3689M}. The QPP patterns may have pronounced amplitude and/or instantaneous oscillation period modulation, and also the signal shape may be highly anharmonic. It means that the oscillation energy is intrinsically distributed over several Fourier frequencies, making the detection techniques based on the Fourier transform not very relevant \citep[e.g.,][]{2022SSRv..218....9A}. 

\citet{2021SSRv..217...66Z} identified at least fifteen mechanisms which could be responsible for the appearance of QPP. These mechanisms include the modulation of the magnetic reconnection rate or parameters of the emitting plasma by magnetohydrodynamic (MHD) waves, and self-induced repetitive reconnection \citep{2018SSRv..214...45M}. As different mechanisms produce QPP with different properties, such as oscillation periods, signal shapes, modulation, etc., one would expect the existence of several different classes of QPP. However, the task of the QPP taxonomy remains an outstanding problem. Obviously, the detection technique must be fine-tuned for the detection of QPP of a specific class. Similarly, the search for statistical relationships among different observables should be conducted within a specific QPP class.

One of the clearly identified class are QPP of co-called SUMER oscillations, named after the instrument used in the first detection of this oscillatory phenomenon \citep[e.g.,][]{2011SSRv..158..397W, 2021SSRv..217...34W}. In spectral observations, SUMER oscillations appear as periodic alternate Doppler shift of a coronal emission line. The oscillation pattern is highly harmonic. A typical period of the SUMER oscillations is about several minutes. The oscillation decays very rapidly, with the exponential damping time being comparable to the oscillation period. SUMER oscillations are interpreted as standing slow magnetoacoustic waves in a coronal loop. A related phenomenon are sloshing oscillations, characterised by a slow magnetoacoustic wave bouncing between footpoints of the loop \citep[e.g.,][]{2019ApJ...884..131R}.  QPP of the SUMER class appear in the decay phase of a flare as, e.g., the modulation of the intensity of the soft X-ray radiation, EUV radiation associated with hot plasma, and radio \citep[e.g.,][]{2012ApJ...756L..36K}. Similar QPP are detected in the soft X-ray radiation produced by stellar flares.
The linear scaling of the damping time with the oscillation period has been established in both solar and stellar flares \citep{2016ApJ...830..110C}.
Furthermore, similar QPP are detected in the white light emission produced by stellar flares \citep[e.g.,][]{Pugh2016, 2023PASP..135f4201B}. 

The search for QPP in solar flares is complicated by a large number of events, which is typically about 10,000 of C-, M-, and X-class flares during an eleven-year cycle, and much more lower-power flares. On the other hand, the large number of events allows for the statistical validation and comparison of various theories, and also creates a ground for the application of machine learning (ML) techniques. ML techniques have been intensively applied to various tasks of modern solar physics for several years. It includes applications to space weather forecasting, including solar flares \citep[e.g.,][]{2013SoPh..283..157A, 2015ApJ...798..135B, 2018ApJ...858..113N, 2019SpWea..17.1166C} and coronal mass ejections \citep[e.g.,][]{2019ApJ...881...15W, 2021JSWSC..11...39G}, and predictors of the geomagnetic activities \citep[e.g.,][]{2009SpWea...7.4004V}; identification of morphological features in prominences, and their relationship with solar activity \citep[e.g.,][]{2024ApJS..272....5Z}; image super-resolution techniques \citep[e.g.,][]{2024SoPh..299...36X}; recognition of coronal loops \citep[e.g.,][]{2024ApJS..270....4W}; inferring transverse velocities at the solar surface \citep[e.g.,][]{2023ApJ...956...83T}; extrapolation of solar global magnetic fields\citep[e.g.,][]{2020ApJ...903L..25J}; and may others. 
Machine learning techniques have also been applied to the detection of stellar flares in photometric surveys \citep[e.g.,][]{2018A&A...616A.163V}, see also \citet{2020WDMKD..10.1349F},
and the search for quasi-periodic eruptions in the X-ray emission from the nuclei of galaxies \citep[e.g.,][]{2023RASTI...2..238W}.
Thus, the application of an ML approach may significantly advance the QPP study too. Promising research avenues are, for example, an ML-based identification of QPP classes, and the application of a pattern-recognition approach to the QPP detection.

The aim of this paper is to develop and test an ML technique for the detection of QPP patterns of the SUMER class, i.e., rapidly-decaying highly harmonic oscillatory patterns in lightcurves, and provide the research community with an open access software package implementing this technique. The paper is organised as follows. In Section~\ref{sec:data} we describe the synthetic data which are used for training a neural network. 
Section~\ref{sec:archit} describes the neural network architecture together with its performance on the synthetic data. In Section~\ref{sec:real_data}, we apply the developed neural network to real flare lightcurves. Then, Section~\ref{sec:install} contains the information about the browser application to use the network. Finally, Section~\ref{sec:conc} concludes the results and outlines the possible next steps to develop the approach presented in this paper.

\section{Synthetic data creation} \label{sec:data}

Usually, a large amount of data is required for training and validating deep learning (DL) models. For our problem, the dataset should contain at least several thousands of lightcurves with and without QPP. Moreover, in training, we need to provide the network with a so-called ground truth knowledge, i.e., a definitive answer as to whether the given flare lightcurve contains a QPP signal or not, which is not always available in the existing flare catalogues.
For this reason, we construct an extensive synthetic dataset of flare lightcurves and rely on it for training and validating our network.
To generate the synthetic dataset, we follow the procedure similar to that presented in \citet{Broomhall2019}, which is briefly outlined below.

Our synthetic flare lightcurves have the following general form,
\begin{equation}
\label{eq:lc}
    \mathcal{I}\left(t\right) = \mathcal{F}\left(t\right) + \mathcal{Q}\left(t\right) + \mathcal{N}\left(t\right).
\end{equation}
Here, $\mathcal{F}(t)$ is the flare profile {with the peak amplitude $A_\mathrm{flare}$}, generated using one of the three flare shape models proposed hitherto, 
\begin{eqnarray}
\label{eq:gryciuk}
    &\displaystyle \mathcal{F}_{1}(t) =A_\mathrm{flare}\frac{\mathrm{G}\left(t\right)}{\mathrm{Max}\left(\mathrm{G}\left(t\right)\right)},\\ 
    &\displaystyle \mathrm{G}\left(t\right)=\exp\left[D\left(B-t\right)+\frac{C^2D^2}{4}\right]\left[\mathrm{erf}\left(Z\right)-\mathrm{erf}\left(Z-\frac{t}{C}\right)\right], \nonumber \\ 
    &Z=\left(2B+C^2D\right)/2C, \nonumber
\end{eqnarray}
based on the convolution of the Gaussian energy deposition and exponential energy dissipation with $B$, $C$, $D$ being free parameters \citep{Gryciuk2017},
\begin{equation} \displaystyle
\label{eq:gaussian}
    \mathcal{F}_2(t) = 
     \left\{
    \begin{array}{l}
     \displaystyle  A_\mathrm{flare} \exp\left(-\frac{t^2}{2\sigma_\mathrm{rise}}\right), \quad t<0 \\
     \displaystyle  A_\mathrm{flare} \exp\left(-\frac{t^2}{2\sigma_\mathrm{decay}}\right), \quad t\geq0
    \end{array}
  \right.
\end{equation}
which is a guessed flare profile approximated by two half-Gaussian curves of different widths {$\sigma_\mathrm{rise}$ and $\sigma_\mathrm{decay}$} \citep{Broomhall2019}, and 
\begin{equation}
\label{eq:davenport}
    \mathcal{F}_3(t) =
      \left\{
    \begin{array}{l}
      A_\mathrm{flare}\left(1+1.941t-0.175t^2-2.246t^3-1.125t^4\right), \quad -1 \leq t < 0\\
      0.948A_\mathrm{flare}\exp\left(-0.965t\right), \quad 0 \leq t < 1.6 \\
      0.322A_\mathrm{flare}\exp\left(-0.290t\right), \quad 1.6 \leq t \leq 19 \\
    \end{array}
  \right.
\end{equation}
derived as an empirical template of a stellar flare in Kepler observations \citep{Davenport2014}. Then, the flare duration is scaled from its initial value of 20 to the new length $L_\mathrm{flare}$ and the flare peak is shifted to the right by a value determined by the \lq\lq shift\rq\rq\ parameter (see Table~\ref{tab:params}). As a final step, the constant offset value is added to the flare.

For the QPP signal $\mathcal{Q}(t)$ in Eq.~(\ref{eq:lc}), in this work we focus on a particular type of rapidly decaying harmonic QPP events in the flare decay phase, proposed by \citet{2019PPCF...61a4024N} and which can be associated with the dynamics of standing slow magnetoacoustic waves, also known as SUMER oscillations, in a flaring loop \citep[e.g.,][]{2021SSRv..217...34W, 2019ApJ...874L...1N},
\begin{equation}
\label{eq:qpp}
    \mathcal{Q}\left(t\right) = A_\mathrm{qpp} \exp\left(-\frac{t}{\tau}\right)\cos\left(\frac{2\pi t}{P}+\phi\right).
\end{equation}
In Eq.~(\ref{eq:qpp}), $A_\mathrm{qpp}$ is the QPP amplitude; $P$, $\tau$, and $\phi$ stand for the QPP period, decay time, and phase, respectively.
The QPP start time $t_\mathrm{start}$ is chosen to be about the time of the flare peak $t_\mathrm{peak}$, but not exactly coinciding with it (see the parameter $t_\mathrm{start} - t_\mathrm{peak}$ in Table~\ref{tab:params}).
Next, the noise component $\mathcal{N}(t)$ is introduced into the lightcurve, which is modelled as either white noise, red noise, or a combination of both types of noise. The  red noise component is generated as
\begin{equation}
\label{eq:noise}
    \mathcal{N}_i = r\mathcal{N}_{i-1}+\sqrt{\left(1-r^2\right)}w_i,
\end{equation}
where $\mathcal{N}_i$ is the red noise value at the $i$-th instant of time, obtained from a white-noise component $w_i$, with the correlation coefficient between successive data points, $r$. The white-noise component $w_i$ is generated from a Gaussian distribution with zero mean and a standard deviation scaled relatively to the flare amplitude.

\begin{figure}[h]
\includegraphics[width=\textwidth]{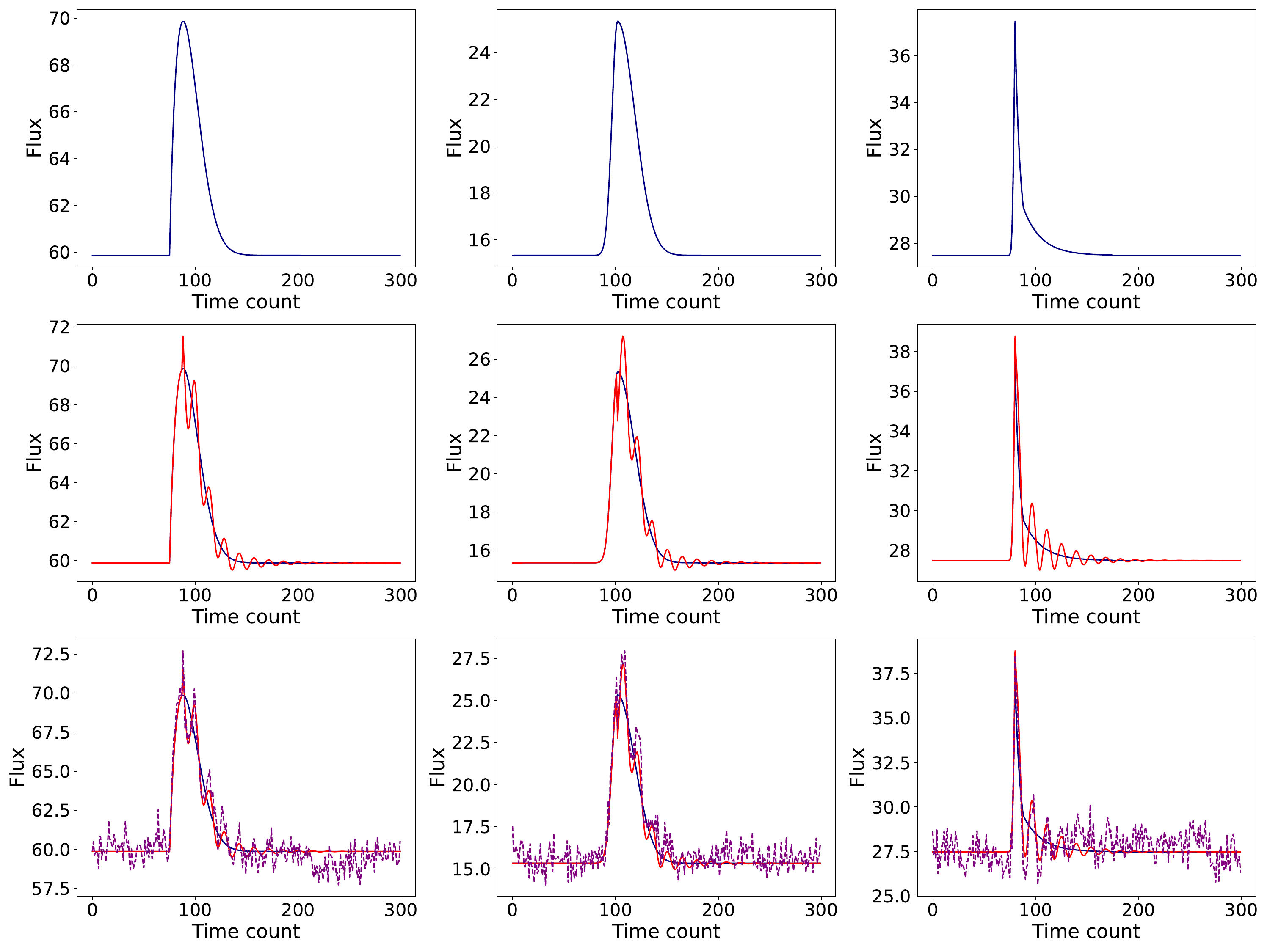}
\caption{Examples of synthetic lightcurves given by flare models Eq.~(\ref{eq:gryciuk}), left, Eq.~(\ref{eq:gaussian}), middle, and Eq.~(\ref{eq:davenport}), right. {The top row:} the initial flare profiles. {The middle row:} the initial flare profiles with added QPP (red curve), see Eq.~(\ref{eq:qpp}). {The bottom row:} the same as in middle row but with added red and white noise (purple dashed curve), see Eq.~(\ref{eq:noise}). \label{fig:data_gen}}
\end{figure}

\begin{figure}[ht]
\label{fig:data_portrait}
\includegraphics[width=18cm]{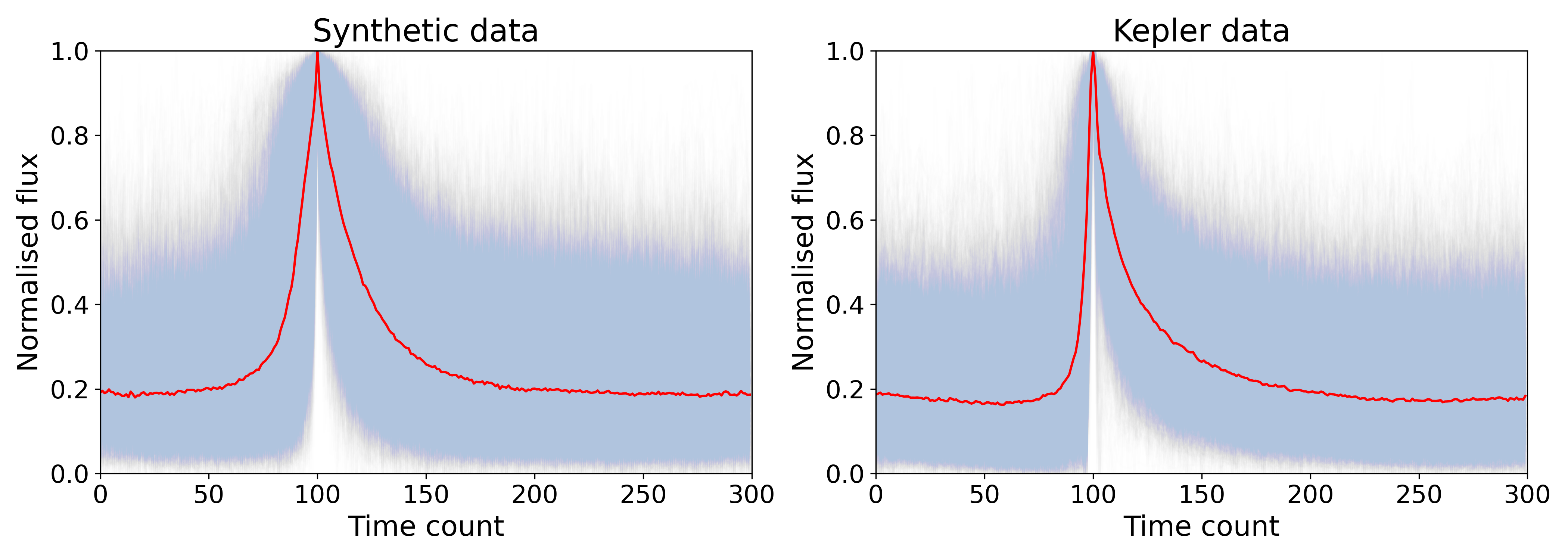}
\caption{Simultaneous plot of 2400 aligned and min-max scaled light curves from the synthetic dataset (left panel) and min-max scaled light curves from stellar flare catalogue, observed with Kepler \citep{Balona2015}.}
\end{figure}

To train the ML model, we created 90,000 synthetic flare lightcurves using Eqs.~(\ref{eq:lc})--(\ref{eq:noise}), in which 50\% of lighcurves have QPP and the other 50\% of lighcurves do not (the QPP amplitude $A_\mathrm{qpp}$ set to zero). Each synthetic flare lightcurve contains 300 data points.  In our dataset, we use nine combinations of the above-mentioned flare shapes $\mathcal{F}_{1,2,3}(t)$ and white/red/white+red noise $\mathcal{N}(t)$. The synthetic flare, QPP and noise parameters we used for the creation of our dataset are summarised in Table~\ref{tab:params}, where $U\left(a, b\right)$ stands for a random value uniformly distributed between $a$ and $b$, and $N\left(\mu, \sigma\right)$ denotes a random value normally distributed with a mean $\mu$ and a standard deviation $\sigma$.
The examples of the flare lightcurves with QPP, generated as described above, are shown in Fig.~\ref{fig:data_gen}, for all three flare profile models $\mathcal{F}_{1,2,3}(t)$.
The left panel of Fig.~\ref{fig:data_portrait} shows {2400} randomly selected synthetic flare lightcurves plotted together, for illustration of the general shape of our dataset. To create this plot, the peaks of the selected lightcurves were positioned at one-third of the time domain and then scaled to unity.

To finalise our synthetic dataset, the created flare signals were randomly shuffled and split into train, test and validation subsets in a proportion of 80\%, 10\% and 10\%, respectively. The synthetic dataset can be accessed online via Harvard Dataverse repository\footnote{\url{doi.org/10.7910/DVN/UNRTN6}}, where one can find all necessary source and data files and description.
\\\\
\begin{table}
\caption{Parameters of synthetic dataset. \label{tab:params}}
\begin{tabular}{ c c c c } 
 \hline
 Parameters & Gryciuk $\mathcal{F}_1(t)$ & Two half-Gaussians $\mathcal{F}_2(t)$ & Davenport $\mathcal{F}_3(t)$\\ 
 \hline
 $L_\mathrm{flare}$ & $U\left(50, 300\right)$ & $U\left(50, 300\right)$ & $U\left(50, 300\right)$ \\ 
 shift & $U\left(0, 300-L_\mathrm{flare}\right)$ & $U\left(0, 300-L_\mathrm{flare}\right)$ & $U\left(0, 300-L_\mathrm{flare}\right)$ \\ 
 $A_\mathrm{flare}$ & $10+N\left(0, 2\right)$ & $10+N\left(0, 2\right)$ & $10+N\left(0, 2\right)$ \\
 $\sigma_\mathrm{rise}$ & n/a & $U\left(1, 3\right)$ & n/a \\
 $\sigma_\mathrm{decay}$ & n/a & $U\left(5, 20\right)$ & n/a\\
 $B$ & $U\left(0, 5\right)$ & n/a &  n/a\\
 $C$ & $U\left(1, 5\right)$ & n/a & n/a\\
 $D$ & $U\left(0.5, 1\right)$ & n/a & n/a\\
 offset & $U\left(0, 100\right)$ & $U\left(0, 100\right)$ & $U\left(0, 100\right)$ \\
 \hline
 $A_\mathrm{qpp}/A_\mathrm{flare}$ & $U\left(0.05, 0.5\right)$ & $U\left(0.05, 0.5\right)$ & $U\left(0.05, 0.5\right)$\\
 $L_\mathrm{flare}/P$ & $U\left(5, 15\right)$ & $U\left(5, 15\right)$ & $U\left(5, 15\right)$ \\
 $\tau/P$ & $U\left(1, 5\right)$  & $U\left(1, 5\right)$ & $U\left(1, 5\right)$ \\
 $\phi$ & $U\left(0, 2\pi\right)$ & $U\left(0, 2\pi\right)$ & $U\left(0, 2\pi\right)$ \\
 $t_\mathrm{start}-t_\mathrm{peak}$ & $ U\left(-0.05L_\mathrm{flare}, 0.1L_\mathrm{flare}\right)$ & $U\left(-0.05L_\mathrm{flare}, 0.1L_\mathrm{flare}\right)$ & $U\left(-0.05L_\mathrm{flare}, 0.1L_\mathrm{flare}\right)$\\
\hline
White S/N & $U\left(1, 5\right)$ & $U\left(1, 5\right)$ & $U\left(1, 5\right)$\\
Red S/N & $U\left(1, 5\right)$ & $U\left(1, 5\right)$ & $U\left(1, 5\right)$ \\
$r$ & $U\left(0.81, 0.99\right)$ & $U\left(0.81, 0.99\right)$ & $U\left(0.81, 0.99\right)$\\
\hline
\end{tabular}
\end{table}

\section{Neural network architecture and performance} \label{sec:archit}
\begin{figure}[h]
\includegraphics[width=\textwidth]{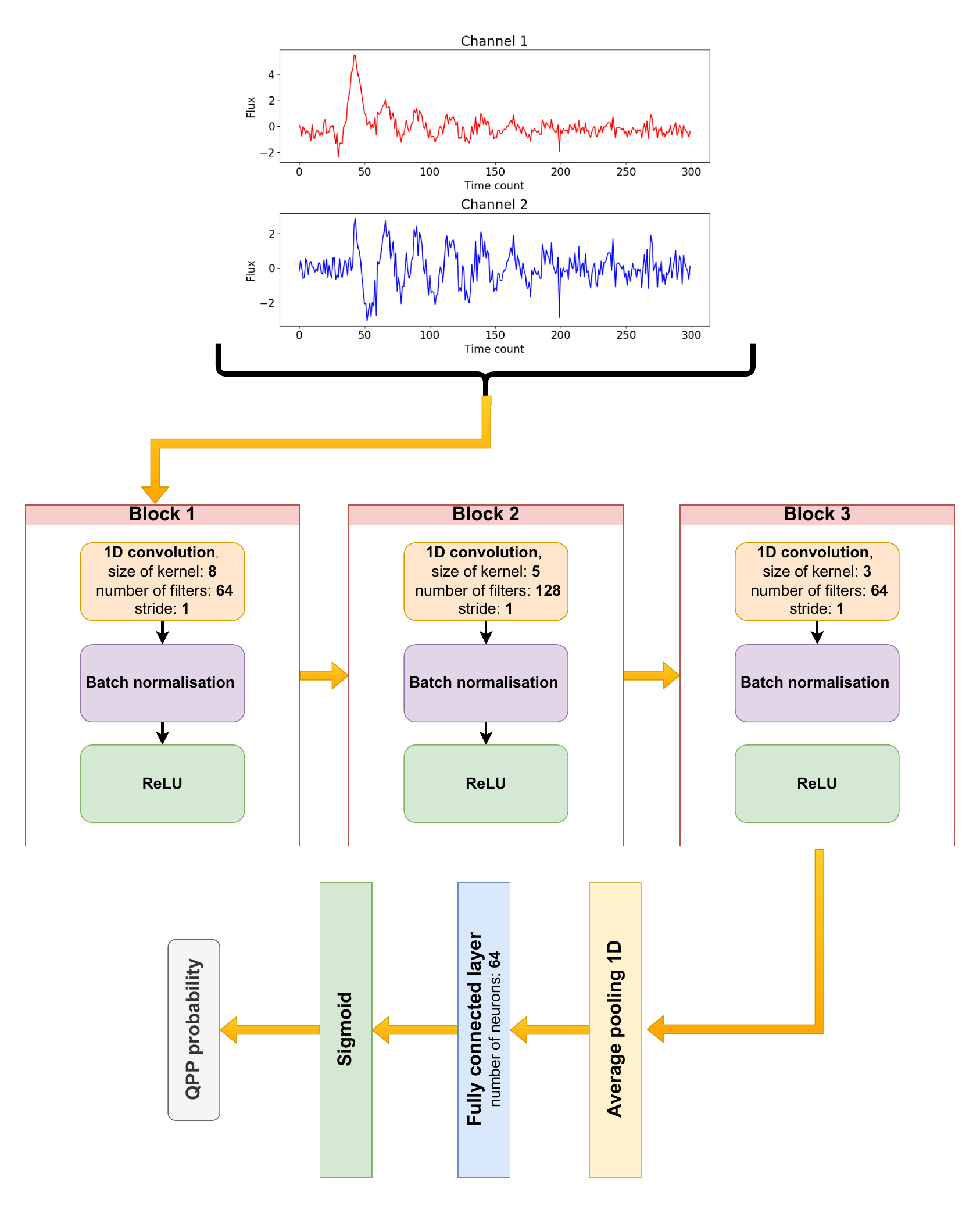}
\caption{Architecture of the FCN for a QPP detection task.}
\label{fig:architecture}
\end{figure}

In this work, we use the Fully Convolutional Network (FCN) architecture proposed by \cite{Wang2017} for a time series classification task. Our PyTorch Lightning implementation is based on the TensorFlow implementation from \citet{IsmailFawaz2019}. The FCN architecture is shown in Fig.~\ref{fig:architecture} and consists of 3 consequent blocks of 1D convolution, batch normalization and Rectified Linear Unit (ReLu) activation function followed by 1D average pooling, one fully connected layer and a sigmoid activation function. As a result, an input times series is transformed into a QPP probability estimation $\mathcal{P}_\mathrm{QPP}$ ranging from 0 to 1. It is usually assumed that if this probability exceeds 0.5 than the ML model finds a positive class.

\begin{figure}[ht]
\includegraphics[width=18cm]{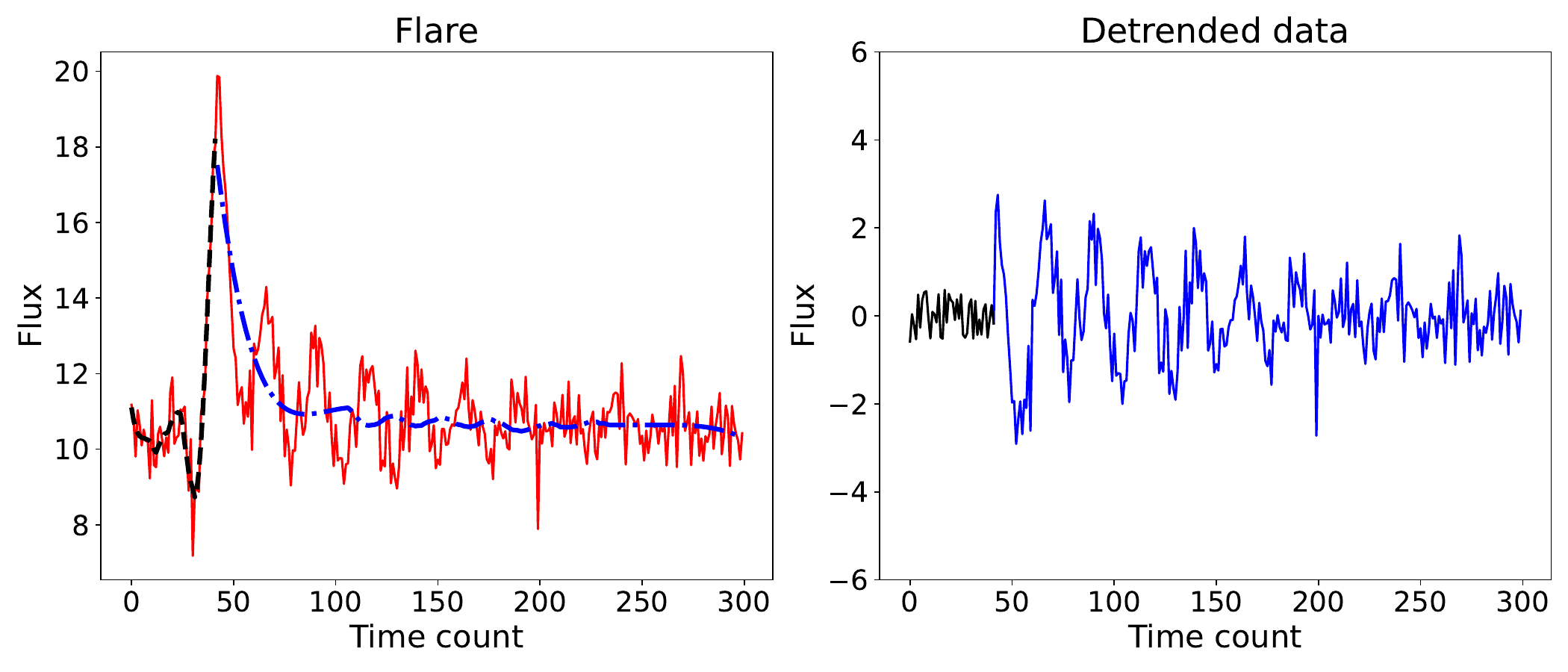}
\caption{Preprocessing of a flare lightcurve to produce two input channels for our FCN (see the architecture in Fig.~\ref{fig:architecture}). {Left panel:} an initial flare time series (red curve) as the input channel 1, with the rise (black dashed curve) and decay (blue dash-dotted curve) trends. {Right panel:} the detrended flare lightcurve as the input channel 2, consisting of the white-noise padded rise phase (black curve) and the decay phase detrended with the Savitzky–-Golay filter (blue curve).
\label{fig:preprocessing}}
\end{figure}

To increase the FCN performance and capability to learn from the data, we added one additional channel to the input data. This channel is produced by detrending the original flare time series. The algorithm of making this channel is as follows: 
\begin{enumerate}
  \item A position of a flare peak is found;
  \item The flare is separated into the rise phase (before the flare peak) and decay phase (after the flare peak);
  \item The decay and rise phase signals are smoothed by the Savitzky--Golay filter of the 4th order and a window width equal to the half-length of the input signal (separately for the rise and decay signals) to produce corresponding trends. For the illustration of smoothing, see the blue dash-dotted (decay phase) and black dashed (rise phase) curves in the left panel of Fig.~\ref{fig:preprocessing};
  \item For the decay phase, the obtained trend is subtracted from the original signal to enhance the visibility of QPP (see the blue curve on the right panel of Fig.~\ref{fig:preprocessing});
  \item For the rise phase, the detrended signal is also obtained, and its standard deviation $\sigma$ is calculated. Then, {the whole} rise phase is padded with white noise produced with $U\left(-\sigma, \sigma\right)$ (see the black curve on the right panel of Fig.~\ref{fig:preprocessing}).
\end{enumerate}

We found that adding the second channel improves the accuracy of the classification by about 5\% on our synthetic dataset by highlighting the QPP features hidden by the trend (i.e. small amplitudes, long periodicities, etc.). It should be mentioned that, in general, this channel is sensitive to the choice of the Savitzky--Golay filter's order and window width \citep[see e.g. the dedicated discussion of this issue in Sec.~4.2 of][]{2020STP.....6a...3K}. Our choice of these parameters allows for obtaining a smooth trend following slow (in comparison with QPP) changes in the lightcurve.
{Moreover, every window filter suffers from the edge effects which were mitigated by interpolating the edges by a polynomial function (the default \lq interp\rq\ mode in the \lq savgol\_filter\rq\ function in the SciPy Python package).}
Nevertheless, the detrended time series is used as an additional complementary channel (in addition to the original time series) in this work, allowing the network to catch the time series properties which may be less pronounced in the original signal. The latter, complementary use of the detrended signal lets us ease the restrictions of the detrending and subjectivity in the choice of the filter's parameters.

Thus, the input data for the FCN consists of two channels. The first channel is the original flare lightcurve (e.g., the left panel of Fig.~\ref{fig:preprocessing}), and the second channel is the flare lightcurve detrended as described above (e.g., the right panel of Fig.~\ref{fig:preprocessing}).
Before feeding into the network, channels in a data sample are standardised independently: 
\begin{equation}
   \hat{\mathcal{I}}_{c,i}^j = \left(\mathcal{I}_{c,i}^j - M\mathcal{I}_{c}^j\right) / \sigma \mathcal{I}_{c}^j,
\end{equation}
where $\mathcal{I}_{c,i}^j$ is the $c$-th data channel of the $j$-th data sample at the $i$-th time count, and $M\mathcal{I}_{c}^j$ and $\sigma \mathcal{I}_{c}^j$ are mean and standard deviation of the time series $\mathcal{I}_{c,i}^j$.

\begin{figure}[h]
\includegraphics[width=\textwidth]{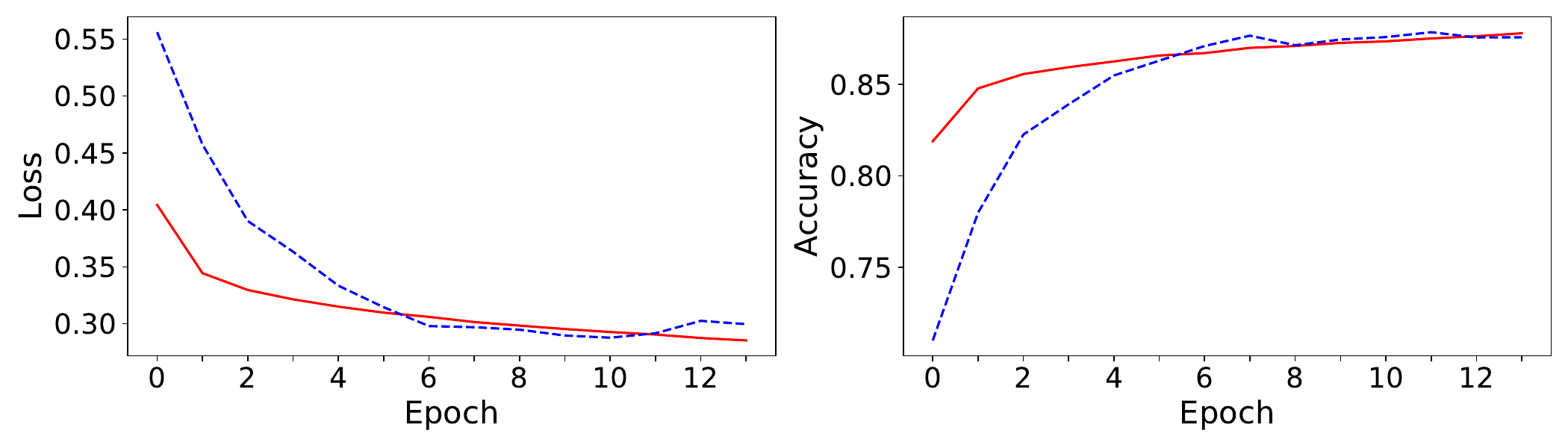}
\caption{Left panel: the dynamics of the loss function during the training process for the train (red curve) and validation (blue dashed curve) datasets. Right panel: the dynamics of the accuracy score on the train (red curve) and validation (blue dashed curve) datasets during the training process.}
\label{fig:train_curves}
\end{figure}

\begin{figure}[h]
\includegraphics[width=\textwidth]{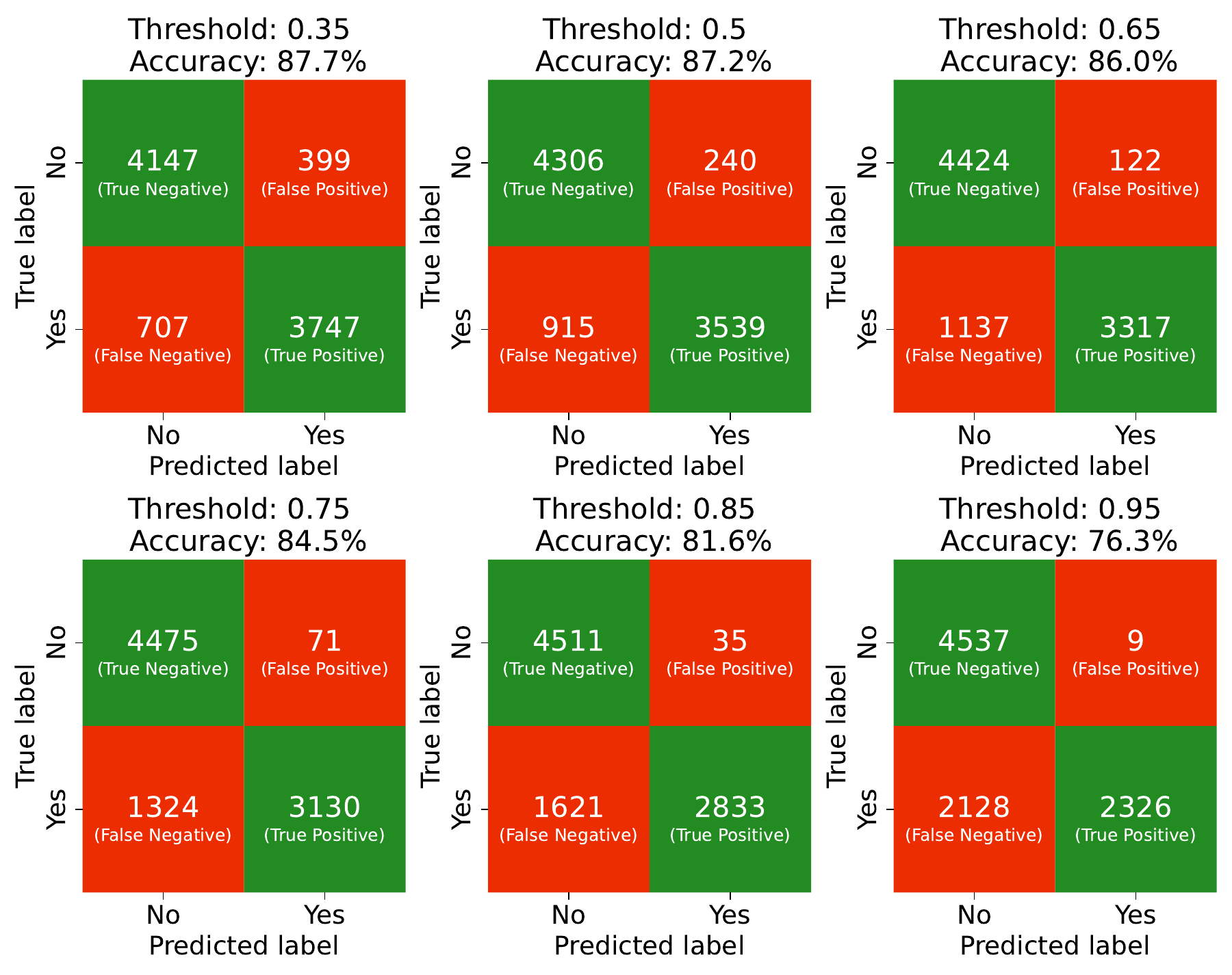}
\caption{Confusion matrices build using the synthetic test dataset for 0.35, 0.5, 0.65, 0.75, 0.85, and 0.95 QPP-detection threshold values.}
\label{fig:cm}
\end{figure}

The FCN has been trained with 64 sample batches on the train dataset using Adam optimizer \citep{Kingma2014} with learning rate of 0.001. The training progress has been tracked using the validation dataset. The early stopping has been used to stop the training process in the validation loss minimum. Figure~\ref{fig:train_curves} shows the dynamics of the loss and accuracy score calculated for the train and validation datasets during the training. It can be seen that the train and validation curves are close to each other, indicating that the FCN has generalized well to new (unseen) data and has not overfitted. Then, the FCN performance has been estimated using the test dataset. The test confusion matrices for different threshold values are shown in Fig.~\ref{fig:cm}. For the 0.5 threshold, the FCN accuracy is $87.2\%$ while precision is $93.6\%$ {(the ratio between the number of true positive answers and a number of all positive answers given by the FCN)}.
For higher thresholds, accuracy decreases (but remains acceptable) with a simultaneous decrease in the number of false positives making the FCN more conservative.

Thus, we can conclude that the FCN performs well on the synthetic dataset and does not experience overfitting  after the training. It indicates the capability of the FCN to generalise the data unseen during the training and makes it potentially applicable to the lightcurves where similar data patterns exist.

\section{Real data examples} \label{sec:real_data}

In this section, we demonstrate the application of our FCN to the existing real data QPP catalogues, observed in solar flares \citep[the Automated Flare Inference of Oscillations (AFINO) catalogue\footnote{\url{aringlis.github.io/AFINO}},][using GOES soft X-ray data]{Inglis2015, Inglis2016} and in stellar flares \citep[][with Kepler in white light]{Balona2015,Pugh2016}.

\subsection{AFINO data}
To create the AFINO catalogue, the Fourier Power Spectral Density (PSD) was calculated for each observed flare time series. Then, the obtained PSD was fitted by three PSD models independently. These models are a single power-law plus a constant (model S0), a single power-law plus a constant and a localized enhancement associated with a QPP signature (model S1), and a broken-power law plus a constant (model S2). For each model, the Bayesian Information Criterion was calculated and the most significant model was chosen.

In our study, we target the rapidly decaying QPP events with periods longer than one minute \citep[motivated by typical observed behaviour of QPP associated with SUMER oscillations in solar and stellar flares, see e.g.,][]{2016ApJ...830..110C}.
In contrast, in the AFINO catalogue, the mean period of QPP events is about  38~s, and only 79 out of 854 events have periods longer than one minute.
Moreover, the QPP events in the AFINO catalogue are sufficiently long-lived, which is required by the PSD-based detection method.
Thus, to look at the most relevant AFINO data, we chose 30 examples from this sub-set of 79 longer-period QPP events, 15 of which have strong preference for model S1 over models S0 and S2 (i.e., QPP detection with high confidence), and the other 15 events have strong preference for model S0 over model S1 and model S2 over model S1 (i.e., no QPP). The lightcurves of the selected flare events were downloaded from the GOES data archive, using GOES SSWIDL software\footnote{\url{https://hesperia.gsfc.nasa.gov/goes/goes.html}}.

For the considered sub-set of AFINO data, our FCN was unable to find any QPPs except one event. This is an expected result because of the above differences between the QPP properties in the AFINO catalogue and in our synthetic dataset. AFINO QPPs are weakly decaying which follows from the imposed PSD model (model S1). This makes it a different QPP type not coinciding with the QPP type considered in our study (long-period and rapidly decaying). In addition, our analysis showed that the QPP amplitudes in the AFINO catalogue are much smaller than those in our synthetic dataset, and AFINO QPPs appear across all flare phases, not only in the decay phase as we assumed in this work.

Thus, the developed FCN cannot be used for AFINO-type QPPs for the reasons mentioned above. However, this may be fixed in future by expanding the synthetic dataset used for the FCN training, to account for the QPP properties from the AFINO catalogue.

\subsection{Kepler data}
\begin{figure}[h]
\includegraphics[width=\textwidth]{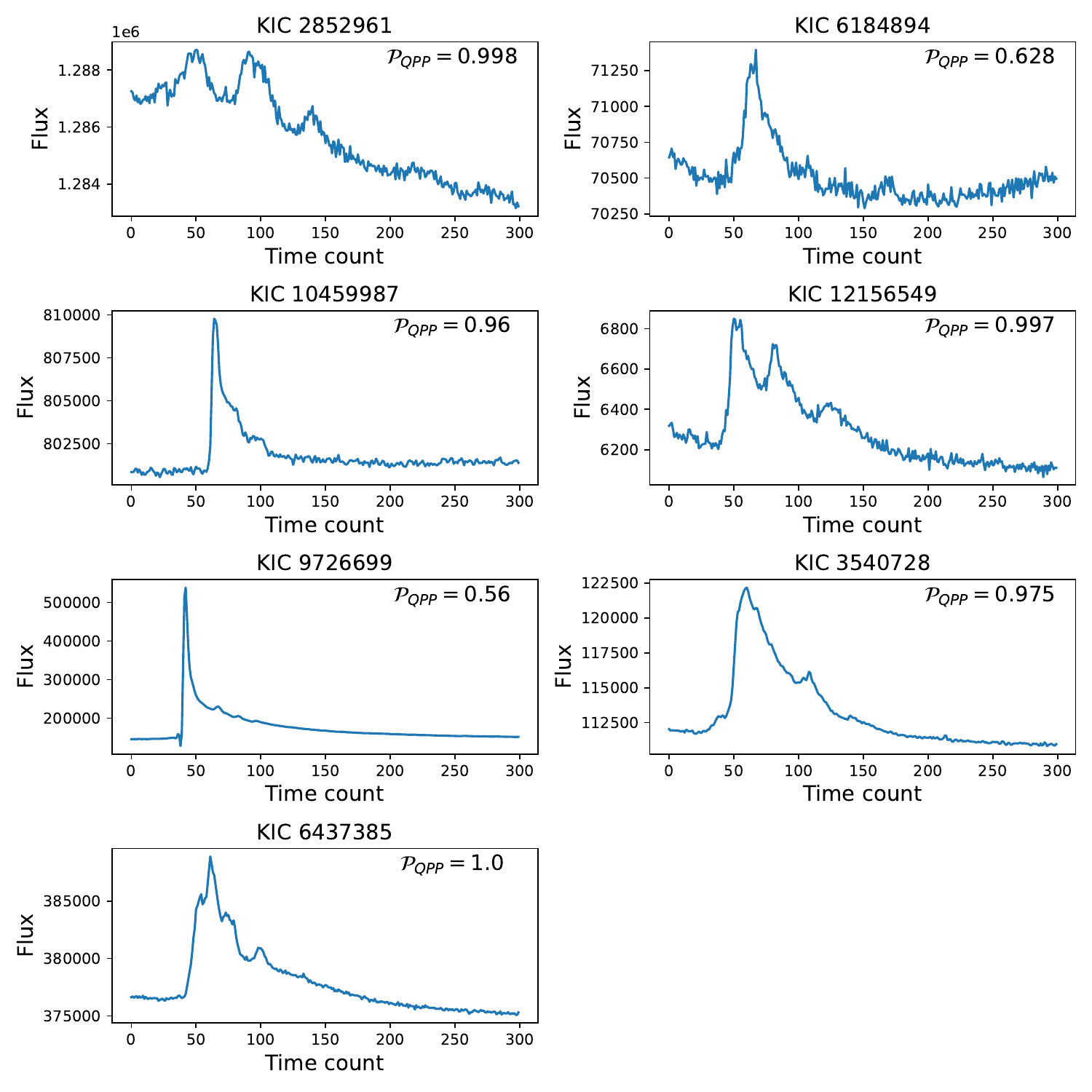}
\caption{Stellar flare lightcurves with QPP from \citep{Pugh2016} where the FCN also found QPP.}
\label{fig:Pugh_yes}
\end{figure}

\begin{figure}[h]
\includegraphics[width=18cm]{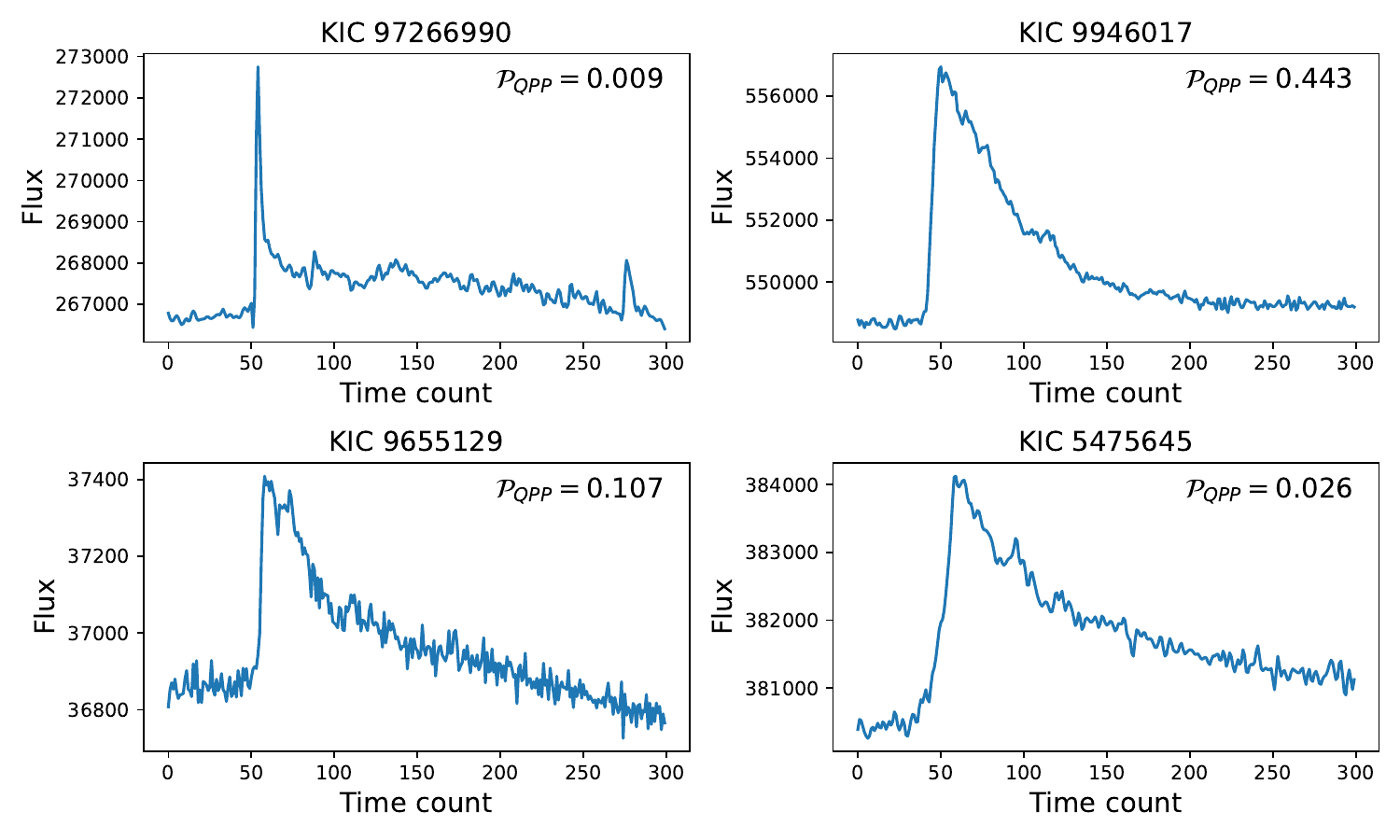}
\caption{Stellar flare lightcurves with QPP from \citep{Pugh2016} where the FCN did not find and QPP.}
\label{fig:Pugh_no}
\end{figure}

In contrast to the AFINO data, QPPs  found in white-light stellar flares look similar to our dataset, even visually. To examine the performance of our FCN on this type of data,  we used the set of 11 QPP events found by \cite{Pugh2016} in Kepler data with 1-min cadence by using {the wavelet power spectrum of autocorrelation function}, which show the stable decaying pattern. Figure~\ref{fig:Pugh_yes} shows 7 out of 11 flare lightcurves from \citep{Pugh2016} where the FCN found QPP, together with the QPP detection probability, $\mathcal{P}_\mathrm{QPP}$. The lightcurves where the FCN did not find QPP are presented in Fig.~\ref{fig:Pugh_no}. This discrepancy with the results of \citep{Pugh2016} seems natural. Indeed, comparing the lightcurves in Figs.~\ref{fig:Pugh_yes} and \ref{fig:Pugh_no}, one may notice that the oscillatory patterns are more obvious (even visually) in Fig.~\ref{fig:Pugh_yes}. In contrast, the lightcurves in Fig.~\ref{fig:Pugh_no} where the FCN was not able to detect QPP in comparison with \citet{Pugh2016} seem to be more noisy and to have smaller relative amplitudes of possible QPP. 

\begin{figure}[h]
\includegraphics[width=18cm]{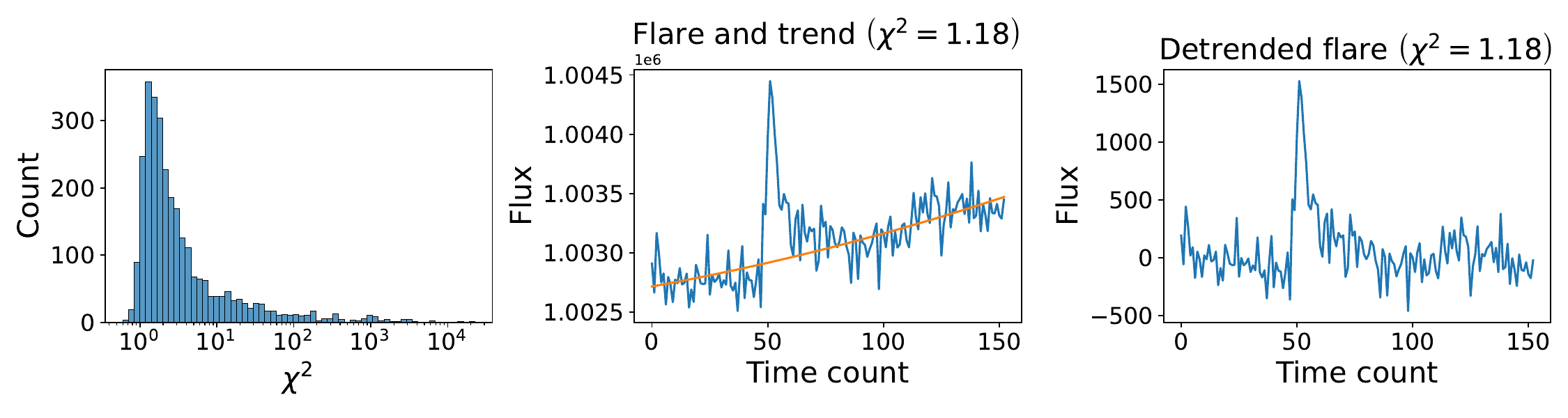}
\caption{{Left panel:} the histogram of the $\chi^2$-criterion characterising how good the slowly-varying background flare trend is fitted by a parabolic function, for stellar flare catalogue of \citet{Balona2015} based on the white-light Kepler data. {Middle panel:} a flare light curve and its parabolic trend. {Right panel:} a flare light curve with the parabolic trend removed.\label{fig:kep_prep}}
\end{figure}

The FCN performance on QPPs found by \cite{Pugh2016} demonstrates that the FCN is able to detect QPPs in real lightcurves and can be used as a data sieve to detect, at least, the most obvious QPPs events. To proceed with this conclusion, we took the data from the stellar flare catalogue \citep{Balona2015} describing 3140 stellar flares observed with Kepler.
The short cadence flare light curves were downloaded by their KIC and flare peak time from the flare catalogue using Lightkurve python library \citep{2018ascl.soft12013L}.
Before using this data as the FCN input, we detrended the lightcurves using a parabolic fit, to get rid of the star rotation effect (see Fig.~\ref{fig:kep_prep}). In this fitting procedure, the second third of the time series where the flares are usually present was masked out so that the fit mainly captures the slowly-varying background trend.
Next, we examined how well this slowly-varying trend best-fits the data using the $\chi^2$-criterion, calculated at the first and last thirds of the time series. The left panel of Fig.~\ref{fig:kep_prep} shows the histogram of the obtained $\chi^2$ values. Using it, we empirically set our $\chi^2$ threshold to 20, which accounts for most of the flare events and filters out the outliers. After filtering the lightcurves with $\chi^2>20$ and removing their slowly-varying background trends (see the middle and right panels of Fig.~\ref{fig:kep_prep}), we selected only flares with peaks exceeding the mean background flux value by 3 standard deviations, that gave us a final set of 2274 flare lightcurves to process with our FCN. The right panel of Fig.~\ref{fig:data_portrait} demonstrates all these selected flare lightcurves plotted together after their min-max normalisation. {It may be noted that we had to remove this slowly-varying background trend because it was not included in our synthetic lightcurves to train the network. However,  in future studies, it can be added to the training dataset to avoid the need for detrending.}

The obtained Kepler lightcurves were used as the FCN input to find QPPs. Table~\ref{tab:freq} presents the fraction of the lightcurves in which the FCN detected QPPs, for different QPP detection threshold values. For example, for $\mathcal{P}_\mathrm{QPP}>0.5$ the fraction of Kepler flares with QPPs is found to be about 23\%. However, for a more conservative $\mathcal{P}_\mathrm{QPP}>0.95$ threshold, the QPP detection rate is about 7\% which is comparable to the recent result of QPP detection in TESS flares \citep{2021SoPh..296..162R} and is approximately factor of two higher than the previous estimation of the QPP detection rate in Kepler flares \citep{Pugh2016, Balona2015}. Furthermore, these 7\% detection rate of QPP in stellar white light flares seem to be very similar to the statistics of QPP in weak solar C-class flares \citep{2020ApJ...895...50H}. Thus, we consider the performance of our FCN in this case very reasonable. For all the lightcurves where the FCN found QPP with $\mathcal{P}_\mathrm{QPP}>0.95$ threshold (159 events in total), the star's KIC identifier and the BJD time of flare peak are summaries in Table~\ref{tab:params_qpp} in Appendix~\ref{sec:app}. This data may be used for a more detailed analysis of QPP, including with other methods, in future.

\begin{table}
\centering
\caption{The fraction of QPP found with the FCN in the Kepler stellar flare catalogue \citep{Balona2015} for different threshold values. A total of 2274 stellar flares were considered.  \label{tab:freq}}
\begin{tabular}{ l c c c c c c c c } 
 \hline
 Threshold & 0.35 & 0.5 & 0.65 & 0.75 & 0.85 & 0.95 & 0.99 \\ 
 QPP fraction, \% & 29.2 & 23.4 & 18.4 & 15.0 & 11.8 & 7.0 & 3.8 \\
\hline
\end{tabular}
\end{table}

\section{FCN installation guide and usage example} \label{sec:install}

The synthetic dataset used in this study and the source files for the developed FCN are available in open access via a GitHub repository\footnote{\url{github.com/BelovSA/QPP-Detection}} and a Harvard Dataverse dataset. The GitHub repository includes two Jupyter notebooks in Notebooks folder (one for synthetic dataset generation and one for the FCN training) and an easy to install and user-friendly Streamlit browser application, created by us for running the developed FCN. To use this application, one should implement the following commands (written in Unix format, assuming Anaconda/Miniconda distribution v.24.1.2 or later has been pre-installed):
\begin{enumerate}
    \item Clone or copy the project repository from GitHub.
    \item In terminal, change your working directory to the root directory of the project:\\ \verb|cd QPP-Detection| (or \verb|cd QPP-Detection-main|).
    \item Create new Anaconda environment (can take several minutes):\\
    \verb|conda env create -f ./Environment/env.yml|
    \item Activate the environment created:\\
    \verb|conda activate qpp_detection|
    \item Run the application\footnote{In case of trouble, the update of \texttt{streamlit} and \texttt{PyTorch-lightning} libraries may require:\\
    \texttt{conda update <library\_name>}}:\\
    \verb|streamlit run ./Application/app.py|
    \item If the steps above are successful you can open the application (if it is not opened automatically) in your browser by the url:\\
    \url{http://localhost:8501} (or a similar one specified in terminal after running the previous command) 
    \item Use the graphical interface to detect QPP. The input should be \verb|.csv| files with the lightcurve data in the column named \lq flux\rq\ (other columns, if any, are ignored). The example lightcurves taken from \citep{Pugh2016} and shown in Fig.~\ref{fig:Pugh_yes} and Fig.~\ref{fig:Pugh_no} can be found in the \lq\lq Test data\rq\rq\ folder in the project directory.
\end{enumerate}
Steps 1--3 should be done once to install the application, while steps 1 and 4--7 should be executed each time. The next subsection illustrates the use-case for the application. For the Windows computers, steps 1--7 should be done via Anaconda Prompt, not Windows shell.

\begin{figure}[h]
\centering
\includegraphics[width=15cm]{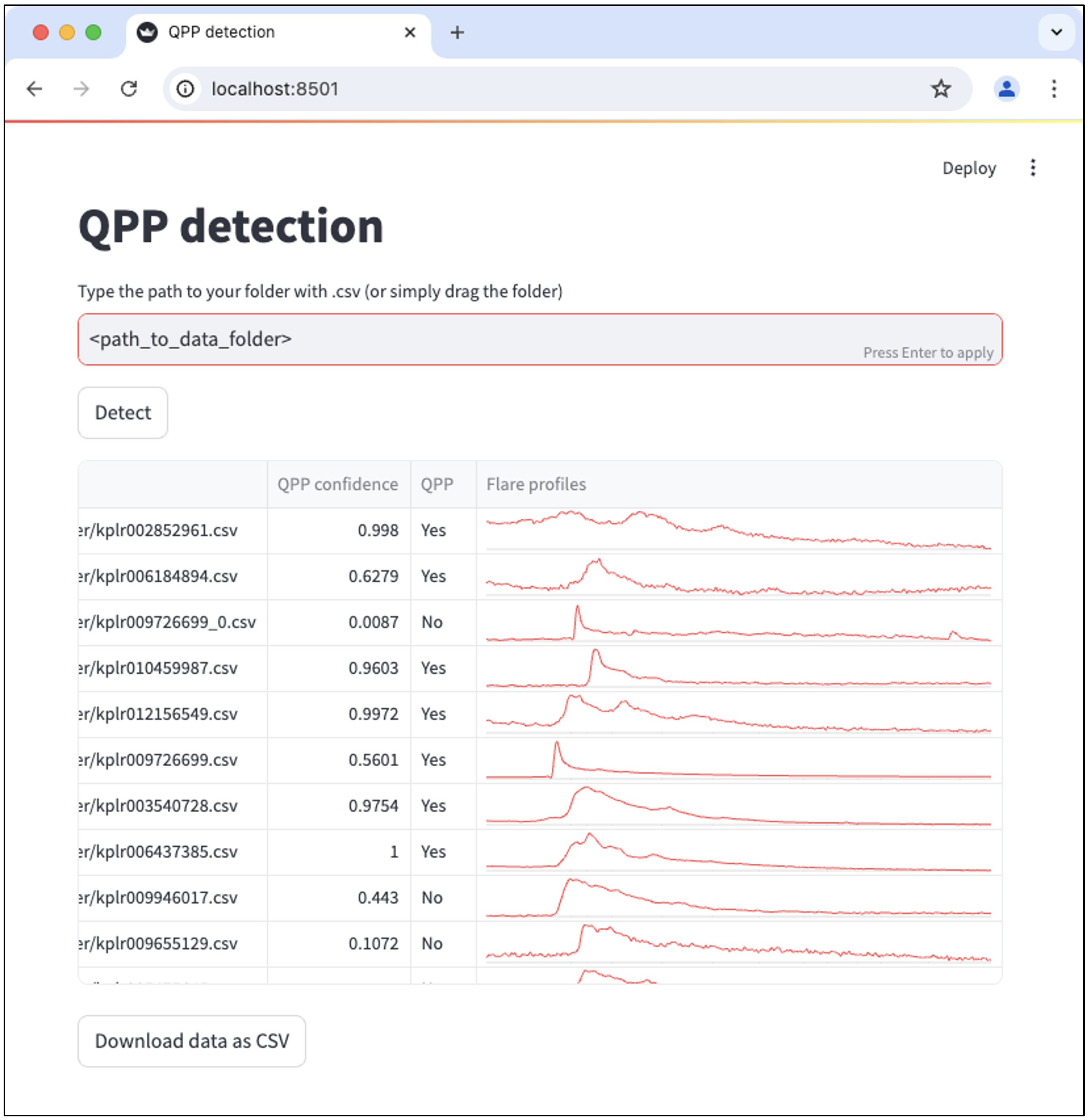}
\caption{The graphical interface of the browser application for the FCN-based QPP detection in the flare lightcurves data.
\label{fig:window}}
\end{figure}

Once steps 4--6 are done, the user should see the following window in a browser as shown in Fig.~\ref{fig:window}. This window initially consists of the search bar, where the folder with the data may be dragged and dropped, and the detect button. After clicking that button, the FCN processes the data and produces the output table containing the QPP probabilities and visualisations of the input flare profiles. Finally, the results may be downloaded by clicking the \lq\lq Download data as CSV\rq\rq\ button for further analysis.

\section{Conclusions and prospects} \label{sec:conc}

In this study we applied the FCN architecture proposed by \cite{Wang2017} to detect QPPs in light curves of solar and stellar flares. For this purpose, we generated 90,000 synthetic lightcurves using the flare models by \citet{Davenport2014}, \citet{Gryciuk2017}, and a guessed flare model made by two half-Gaussians \citep{Broomhall2019}. Then, decaying QPPs were added to the decay phase of 50\% of the lightcurves generated. Also, noise (white, red or both) was added to all the lightcurves. The obtained dataset was divided into train, validation and test datasets.

To train the FCN, we used two input channels: an original lightcurve and a detrended lightcurve with a rise phase padded by white noise.  After training, the FCN showed an accuracy of 87.2\% on the test data with a precision score of 93.6\%. The change of the detection threshold from 0.5 to 0.95 led to a decrease in a number of false positive detections, as well as, to a decrease in the accuracy score making the FCN more conservative.

After testing the FCN performance on the synthetic data, we applied the FCN to real data examples. For the 30 selected samples from the AFINO catalogue \citep{Inglis2015, Inglis2016} 50\% of which contain QPP and 50\% do not, the FCN was able to find only one QPP. The reason for this is that the type of QPP in the AFINO catalogue is apparently different from that in our training dataset.  To test the FCN on data similar to the training data, we used 11 lightcurves of stellar flares with strong evidence of decaying QPP presence found by \cite{Pugh2016} in Kepler observations. For this set, the FCN detected QPPs in 7 out of 11 flares. The remaining 4 lightcurves had either noisy profiles or small QPP amplitudes.

To proceed with Kepler data, we used the information about 3140 lightcurves from the stellar flare catalogue created by \cite{Balona2015}. We detrended these lightcurves using a slowly-varying parabolic fit and filtered out the lightcurves where the fit was unsuccessful using the $\chi^2$-criterion. Additionally, we selected the samples where the peak amplitude exceeded the mean by $3\sigma$. After this preparation, 2274 lightcurves were fed to the FCN. For the threshold $\mathcal{P}_\mathrm{QPP}>0.95$, the FCN detected QPPs in 7\% of all flare lightcurves considered. This is comparable with the QPP detection rate in stellar flares, estimated recently by \citet{2021SoPh..296..162R}.

We made our synthetic dataset and the source code available in open access via Harvard Dataverse and GitHub. Our GitHub repository has two Jupyter notebooks to generate the dataset and train the FCN with it. It also contains the Streamlit browser application, developed by us, which allows one to run the pre-trained FCN and use it for QPP detection in future studies. We also provided a detailed user guide for the installation, running, and using the developed FCN.

We used the FCN time series classifier, based on the architecture proposed in \citep{Wang2017}, as the first step in applying Deep Learning techniques for the QPP detection task. 
As for the next step, a more complicated network architecture can be used including different approaches such as  residual (ResNet) and recurrent (RNN) networks. However, this approach demands powerful hardware and may limit the inference speed that can be crucial for large-scale surveys. Another possible development is data engineering. The additional features may be used as the new FCN input channels. For example, the autocorrelation function can be calculated for the detrended lightcurve and passed to the FCN as an additional channel. Finally, the raw lightcurves can be transformed into 2D spectra (via, for example, the wavelet transform) and then examined by the image recognition techniques for the presence of characteristic QPP signatures.

\begin{acknowledgments}
The work is funded by the STFC Grant ST/X000915/1. DYK and VMN also acknowledge funding from the Latvian Council of Science Project No. lzp2022/1-0017.
\end{acknowledgments}

\software{NumPy, a Python packge for fundamental scientific computing \citep{harris2020array};
SciPy, a Python package for fundamental algorithms for scientific computing \citep{2020SciPy-NMeth};
Lightkurve, a Python package for Kepler and TESS data analysis \citep{2018ascl.soft12013L};
Astopy, a Python package for astrophysical purposes \citep{2013A&A...558A..33A, 2018AJ....156..123A, 2022ApJ...935..167A};
PyTorch Lightning, the deep learning framework;
Streamlit, a Python library for deploying ML projects.}

%






\appendix

\section{Kepler flare with QPP}
\label{sec:app}

\begin{table}
\caption{{KIC and BJD flare peak times for} 159 flares with $\mathcal{P}_\mathrm{QPP}\geq0.95$ found in the flare catalogue \citep{Balona2015}.}
\begin{tabular}{|cc|cc|cc|cc|} 
 \hline
 KIC & BJD & KIC & BJD & KIC & BJD &  KIC & BJD\\ 
\hline
 11560431 & 2456115.361 & 10160534 & 2455064.852 & 11560431 & 2456173.104 & 10355856 & 2455387.912 \\
3430868 & 2455038.978 & 5357275 & 2455119.591 & 11551430 & 2455013.017 & 11551430 & 2455029.862 \\
3239945 & 2456170.505 & 5733906 & 2455099.532 & 4671547 & 2455073.743 & 11560431 & 2456056.445 \\
11560431 & 2456072.751 & 11610797 & 2454981.631 & 11551430 & 2456131.799 & 11665620 & 2455806.894 \\
11548140 & 2455386.295 & 11551692 & 2456170.504 & 5475645 & 2456330.43 & 6106152 & 2455024.179 \\
5357275 & 2455106.976 & 3441906 & 2455952.2 & 9349698 & 2456354.714 & 11709006 & 2455408.873 \\
9895004 & 2455437.151 & 11551692 & 2456042.642 & 10459987 & 2456179.644 & 8429280 & 2455004.753 \\
5557932 & 2455038.978 & 11189959 & 2455114.891 & 11560431 & 2456110.758 & 11560431 & 2456197.291 \\
6106152 & 2455018.608 & 9652680 & 2455090.13 & 1161345 & 2456170.506 & 10063343 & 2455151.39 \\
11560431 & 2456172.309 & 11231334 & 2455491.728 & 5108214 & 2455837.489 & 4273689 & 2455249.467 \\
11560431 & 2455020.996 & 11709006 & 2454971.877 & 4758595 & 2456233.845 & 11551430 & 2455024.522 \\
6548447 & 2455323.093 & 7940546 & 2455768.88 & 11551430 & 2456194.753 & 3239945 & 2456156.116 \\
4543412 & 2455163.976 & 11560431 & 2456117.885 & 4831454 & 2455187.025 & 9761199 & 2455799.706 \\
8651471 & 2455845.632 & 7339343 & 2455005.364 & 7206837 & 2456084.741 & 11551430 & 2456279.936 \\
11709006 & 2455430.144 & 9641031 & 2456219.273 & 11560431 & 2456064.381 & 8429280 & 2455032.148 \\
8429280 & 2455006.133 & 11560447 & 2455055.708 & 11709006 & 2454971.707 & 5609753 & 2456399.539 \\
11560431 & 2455013.574 & 10528093 & 2456262.771 & 4758595 & 2456226.538 & 7841024 & 2454968.701 \\
11548140 & 2455486.206 & 9821078 & 2455623.573 & 11560431 & 2455023.444 & 12644769 & 2456170.501 \\
11551692 & 2456289.977 & 9641031 & 2455022.582 & 11548140 & 2455488.308 & 11665620 & 2455791.997 \\
2300039 & 2455773.901 & 2162635 & 2455806.895 & 12156549 & 2455346.612 & 11560431 & 2456057.61 \\
12156549 & 2455342.383 & 9641031 & 2456157.95 & 7940546 & 2455779.026 & 11560431 & 2456165.778 \\
4758595 & 2456210.581 & 12156549 & 2455376.074 & 11551430 & 2456245.555 & 2302548 & 2456170.505 \\
7940546 & 2455871.277 & 4568729 & 2455056.284 & 9705459 & 2456371.072 & 9641031 & 2456109.318 \\
1025986 & 2455133.355 & 11560431 & 2456117.449 & 9655129 & 2456148.63 & 9655129 & 2456112.973 \\
4671547 & 2455087.968 & 11560431 & 2456150.678 & 11709006 & 2454969.451 & 11551430 & 2456217.422 \\
11560431 & 2456184.967 & 5557932 & 2455039.036 & 4671547 & 2455074.027 & 12156549 & 2455347.197 \\
5522786 & 2456011.854 & 11560431 & 2456070.227 & 12102573 & 2455072.339 & 11560431 & 2455003.17 \\
4939265 & 2456281.794 & 4671547 & 2455090.038 & 6286925 & 2455038.42 & 9641031 & 2456100.371 \\
2300039 & 2455790.849 & 3430868 & 2455039.036 & 7765135 & 2455072.194 & 11551430 & 2456170.457 \\
1161345 & 2455806.895 & 5522786 & 2455871.3 & 11560431 & 2456194.578 & 12156549 & 2455318.629 \\
11551430 & 2456296.8 & 1871056 & 2456156.115 & 6106152 & 2455013.524 & 7692454 & 2456266.216 \\
8429280 & 2455032.246 & 10355856 & 2454982.87 & 9821078 & 2455705.652 & 7940546 & 2456094.427 \\
9833666 & 2456394.052 & 6442183 & 2455860.26 & 11231334 & 2456178.25 & 6205460 & 2455334.886 \\
11560431 & 2455010.158 & 7940546 & 2456259.122 & 12102573 & 2455079.331 & 11551430 & 2456235.537 \\
7206837 & 2455806.894 & 11560431 & 2456186.81 & 11560431 & 2456170.503 & 11560431 & 2456114.382 \\
12156549 & 2455086.645 & 9833666 & 2456423.294 & 11560431 & 2456156.548 & 7940546 & 2455048.776 \\
11560431 & 2456112.473 & 11560431 & 2456142.847 & 7940546 & 2455503.213 & 6106152 & 2455033.037 \\
3128488 & 2454964.786 & 12156549 & 2455067.352 & 4273689 & 2455263.655 & 11709006 & 2454966.828 \\
11231334 & 2455435.206 & 12156549 & 2455374.593 & 11548140 & 2455453.673 & 11551430 & 2456278.089 \\
12418816 & 2455257.868 & 11560431 & 2456132.526 & 4758595 & 2456219.141 &  &  \\
\hline

\end{tabular}
\label{tab:params_qpp}
\end{table}


\bibliography{qpp_fcn}{}
\bibliographystyle{aasjournal}



\end{document}